\documentstyle[aps,prb]{revtex} \def\narrowtext{} \tighten \twocolumn
\begin{document}
\draft
\title{Polarization Selection Rules and Superconducting Gap Anisotropy in
$Bi_2Sr_2CaCu_2O_8$}
\author{M. R. Norman and M. Randeria}
\address{Materials Science Division,
Argonne National Laboratory, Argonne, Illinois  60439}
\author{H. Ding, J. C. Campuzano, and A. F. Bellman}
\address{Materials Science Division,
Argonne National Laboratory, Argonne, Illinois  60439
and Department of Physics, University of Illinois at Chicago,
Chicago, Illinois  60607}

\address{%
\begin{minipage}[t]{6.0in}
\begin{abstract}
We discuss polarization selection rules for angle-resolved photoemission
spectroscopy in Bi2212.  Using these we show that the ``hump'' in the
superconducting gap observed in the $X$ quadrant in our earlier work is
not on the main $CuO_2$ band, but rather on an umklapp band arising from
the structural superlattice.  The intrinsic gap is most likely quite
small over a range of $\pm 10^\circ$ about the diagonal directions.
\typeout{polish abstract}
\end{abstract}
\end{minipage}}

\maketitle
\narrowtext

Recently, our group, by use of angle-resolved photoemission spectroscopy
(ARPES) on high quality single crystals, has found that the superconducting
gap in Bi2212 has a
non-trivial momentum dependence.\cite{ding} In particular, the gap is
suppressed
over a range of $\pm 10^\circ$ about the diagonal directions (at $45^\circ$ to
the CuO bond directions) of the square
lattice zone (for Bi2212, these directions are $\Gamma X$ and $\Gamma Y$
where $X \equiv (\pi,-\pi)$ and $Y \equiv (\pi,\pi)$).  In addition, though,
we found that the gap was actually non-zero over this range, with a weak
maximum along the diagonal and two nodes $10^\circ$ on either side (i.e., a
``hump'').  The simplest
interpretation of such data is that the order parameter has anisotropic
s-wave symmetry.\cite{norman}  Such an order parameter would not be in conflict
with experiments which only probe the existence of nodes, but it would be in
conflict with phase coherence experiments on YBCO which reveal a sign change of
the order parameter under rotation by $90^\circ$,\cite{phase} although
similar experiments have not been performed on Bi2212.

A striking point about our data was that the ``hump'' in the gap along the
diagonal was less pronounced in the $Y$ quadrant of the zone (Fig.~1) than the
$X$ quadrant (Fig.~2).  This led to a potential concern about the
interpretation
of the data because of the presence of an incommensurate superlattice in Bi2212
with a $\vec{Q}$ of $(0.21\pi,0.21\pi)$ which makes the $X$ and $Y$
quadrants inequivalent.\cite{with}  This superlattice
leads to the presence of $\pm\vec{Q}$ umklapp bands in the ARPES
data.\cite{ding,aebi} Because $\vec{Q}$ is along $\Gamma Y$, these umklapp
bands
are well separated from the main band near the diagonal direction in the
$Y$ quadrant, but since $\vec{Q}$ is perpendicular to $\Gamma X$, the two
umklapp bands are predicted to be quite close to the main band near the
diagonal
direction in the $X$ quadrant, and thus should be hard to
resolve (Fig.~3).\cite{ding}  In fact, the main band and superlattice band are
predicted to cross at exactly the same location as where the nodes in the gap
were observed in the $X$ quadrant (Fig.~3).  This suggested the possibility
that the ``hump'' in the gap in the $X$ quadrant was a consequence
of being on the superlattice band instead of the main band.  In an earlier
paper of ours,\cite{norman} we demonstrated that even if the main band had a
gap with d-wave symmetry, the observed gap in the $X$ quadrant could indeed
have this ``hump'', simply because the superlattice band along the diagonal
direction corresponds to a location on the main band about $20^\circ$ away.
There were two problems with such an interpretation.  The first was
why only the superlattice lattice band was observed near the diagonal
in the $X$ quadrant and not the main band.  The second was that a similar
(but smaller) ``hump'' was observed in the $Y$ quadrant, where these arguments
do not apply (that is, the observed gap was definitely on the main band, as
seen in Fig.~3).

As it turns out, the resolution of the first problem was essentially present
in the original data and not realized at the time.  We note that the data in
the $X$ quadrant were taken with a polarization vector along the $\Gamma X$
direction.\cite{ding}  Now, $\Gamma X$ is a symmetry
line of the zone (this direction is reciprocal to the orthorhombic a axis of
Bi2212), and therefore, bands must be even or odd relative to reflections
about this line.  As the band contains copper states of $d_{x^2-y^2}$
symmetry, the main band must be odd relative to this line.  Since the
polarization vector is along this line, then the dipole matrix element
vanishes and therefore emission from the main band cannot be observed.
Now, consider the two superlattice bands which cross each other along
$\Gamma X$ (Fig.~3).  One can make a linear combination of these two bands at
the crossing point. The combination
$\psi(\vec{k}+\vec{Q})+\psi(\vec{k}-\vec{Q})$
has odd symmetry, like the main band, and thus should not be observed either.
The combination $\psi(\vec{k}+\vec{Q})-\psi(\vec{k}-\vec{Q})$,
though, has even symmetry, and is maximally enhanced in this
polarization geometry.  Therefore, the observed gap near the diagonal in
the $X$ quadrant must be on the even symmetry combination of the superlattice
bands, and thus, as conjectured in the previous paragraph, the ``hump'' in
the gap is a superlattice effect.

To explicitly demonstrate this, we have looked at two models for the gap
function.  The first is the standard d-wave gap $d \equiv
cos(k_x+q)-cos(k_y+q)$
where $q$ is $0$ for the main and $\pm 0.21\pi$ for the superlattice bands.
The other is $max(|d|-c,0)$ where $c$ is a constant, which was designed to
have an angular range with zero gap.
The data are then fit assuming the four or six data points nearest the diagonal
are on the superlattice band.  The resulting gaps are plotted along the main
and superlattice sheets in Fig.~2 (note from Fig.~3 that
data for angles less than $35^\circ$ are definitely on the main band).
The comparison to experiment is quite reasonable, especially in the case of
the $|d|-c$ gap (its RMS error in the $X$ quadrant is less than half that of
the d-wave gap\cite{rms}).
The same also applies in the $Y$ quadrant as seen in Fig.~1.
In fact, both fits have a smaller RMS error in the $Y$ quadrant than
the $cos(k_x)cos(k_y)$ function previously used to describe a
``hump'' in the gap.\cite{norman}  We also note that the angular range of zero
gap for the $|d|-c$ model is compatible with NMR data on samples from the same
source as ours, which reveal the presensce of a residual density of states
about 24\% of the normal state value.\cite{nmr}

The observation that the ``hump'' in the $X$ quadrant is a superlattice effect
also explains a number of other puzzling results.  First,
the two spectra nearest $\Gamma X$ had leading edge slopes which were not
resolution limited.\cite{ding}  This can be
understood now since the $\vec{k}$ vectors at which this data were taken
corresponded to an energy near the Fermi energy of the main band but below
the Fermi energy of the superlattice bands (Fig.~3).
Second, previous polarization data\cite{review} indicated a strong polarization
in the $Y$ quadrant but a weak one in the $X$ quadrant.  In the $Y$ quadrant,
both main and superlattice bands are odd relative to $\Gamma Y$ (since
$\vec{Q}$
is along $\Gamma Y$) and thus a strong polarization dependence is indeed
expected.  In the $X$ quadrant, though, when the polarization is perpendicular
to $\Gamma X$, the main band and odd symmetry combination of the superlattice
bands should be observed, whereas when the polarization is parallel to
$\Gamma X$, only the even symmetry combination of the superlattice bands should
be observed.  Our data in the normal state\cite{normal} indicate that the
odd symmetry combination is not observed, perhaps due to final state effects.
Therefore, if the main and even symmetry combination of the superlattice
bands have comparable intensities, then a weak polarization dependence should
be observed in the $X$ quadrant.

This now leaves the remaining problem to be resolved:  the observation of
a ``hump'' in the $Y$ quadrant.  This ``hump'' was based on the two data points
nearest $\Gamma Y$ (Fig.~1).
We note, though, that the error bars of Ref.~\onlinecite{ding}
are consistent with a zero gap for the data point along $\Gamma Y$.
We have gone back and reanalyzed the data for the other point and still
conclude that the
optimal fit has a gap of order 5 meV, and thus not consistent with zero.
We are hestitant, though, to attach too much significance to data at a single
point, and therefore conclude that the current data set is not extensive
enough to definitively prove the existence of a ``hump'' in the $Y$ quadrant.
We caution that the data in the $Y$ quadrant were taken in steps of
$\sqrt{2}^\circ$ in photoemission coordinates, which along the diagonal
corresponds to an energy change (from fits to the dispersion) of over 80 meV
per step.  Although the momentum windows of the data points do overlap with
one another (since the window has a radius of $1^\circ$), the resolution
normal to the Fermi surface is still not as high as one would like.  In the
future, it would be desirable to perform experiments using smaller step sizes
and higher momentum resolution.
Second, the data fits were based on a spectral function which only has
resolution broadening with other effects being absorbed into a model
background function.  The actual spectral function has a strongly energy
dependent lineshape due to self-energy effects.  In the superconducting
state, though, the line broadening will be suppressed for frequencies below
3$\Delta$  if it is due to electron-electron scattering,\cite{coffey}
and such a model appears to be consistent with our data since the
slopes of the leading edge are resolution limited, even along the diagonal.
Since it is a fit to the leading edge which determines the gap, then ignoring
the detailed energy dependence of the broadening should be valid, but one
always worries that a more sophisticated model of the lineshape taking into
account its frequency dependence (which would necessitate using a different
background function) might yield different estimates of the gap.

Given the above, the existence of a ``hump'' in the intrinsic gap is
questionable.  While a ``hump'' necessarily implied anisotropic s-wave
pairing, an extended region of zero or small gap could be consistent with
a number of models.  One of these might be dirty d-wave, with the gap
suppressed in a large angular range about the diagonal due to impurity
scattering.  The NMR data previously mentioned were interpreted in terms of
this
model.\cite{nmr}  There is a problem with such a scenario.
Gaplessness due to dirt shows up in the spectral function as a low
frequency tail.\cite{r1}  In ARPES, though, the gap is determined by measuring
the shift in the leading edge of the spectrum, and this shift
should be relatively unaffected for small concentrations of
impurities.\cite{r2}
If we assume, then, that the suppressed gap region is not due to dirt, then
another possibility is a d-wave gap, but not of the simple $cos(k_x)-cos(k_y)$
form.\cite{rms}  Since this simple form will occur for short-range
interactions,
the implication then is that the interaction must be long-range to obtain a
gap similar to what is observed.
The same argument would also apply to anisotropic s-wave
models (thus, phase coherence experiments would be necessary to determine the
actual symmetry of the order parameter).  In fact, we know of two such models
which can give gaps of this form:  the interlayer
tunneling model of Chakrarvarty et al\cite{chak} and the
poorly screened electron-phonon model of Abrikosov.\cite{alex}  Both of these
models have the property that they are local in $\vec{k}$ space and thus
long-range in real space.

In conclusion, a reanalysis of our data using polarization selection rules
has led us to conclude that the actual gap in Bi2212 is suppressed over a
$\pm 10^\circ$ range about the diagonal, although more experiments at better
momentum resolution will be necessary to completely resolve this issue.
Such a suppressed gap region could
be either due to dirt or to a pairing interaction long-range in space.
Further experiments on samples with controlled defect densities could be used
to differentiate between these two scenarios.

\acknowledgments

We would like to thank Phil Anderson and Roland Fehrenbacher for
useful discussions.
This work was supported by the U.~S.~Department of Energy,
Basic Energy Sciences, under Contract \#W-31-109-ENG-38.

\begin{figure}
\caption{Gap in meV in the $Y$ quadrant versus angle on the Fermi surface
(filled circles) with fits to the data using (a) a d-wave gap and (b) a
$|d|-c$ gap (open circles).}
\label{fig1}
\end{figure}

\begin{figure}
\caption{Gap in meV in the $X$ quadrant versus angle on the Fermi surface
(filled circles) with fits to the data using (a) a d-wave gap and (b) a $|d|-c$
gap (open circles for the main band, crosses for the superlattice (SL) band
nearest the main band).}
\label{fig2}
\end{figure}

\begin{figure}
\caption{Fermi surface from a tight binding fit to the energy dispersions
in the $X$ and $Y$ quadrants of the zone.  The thick line is the main band,
the thin lines the superlattice bands (marked +Q and -Q).  The data points
at which the gap was measured are shown as open circles.}
\label{fig3}
\end{figure}


\begin{references}

\bibitem{ding} H. Ding et al., Phys. Rev. Lett. {\bf 74}, 2784 (1995).

\bibitem{norman} M. R. Norman et al., Phys. Rev. B {\bf 52}, 615 (1995).

\bibitem{phase} D. A. Wollman et al., Phys. Rev. Lett. {\bf 71}, 2134 (1993)
and {\bf 74}, 797 (1995); C. C. Tsuei et al., Phys. Rev. Lett. {\bf 73}, 539
(1994); J. R. Kirtley et al., Nature {\bf 373}, 225 (1995).

\bibitem{with} R. L. Withers et al., J. Phys. C {\bf 21}, 6067 (1988).

\bibitem{aebi} J. Osterwalder et al., Appl. Phys. A {\bf 60}, 247 (1995).

\bibitem{rms}  A gap which is the square of the d-wave gap has the lowest RMS
error of any single function fit we have looked at, with an error only slightly
larger than that of the $|d|-c$ gap (the latter having two fit parameters).
A gap of this type has been suggested by P. W. Anderson.

\bibitem{nmr} K. Ishida et al., J. Phys. Soc. Japan {\bf 63}, 1104 (1994).

\bibitem{review} For a review, see Z.-X. Shen and D. S. Dessau, Phys. Rep.
{\bf 253}, 1 (1995).

\bibitem{normal} H. Ding et al., unpublished.

\bibitem{coffey} C. M. Varma and P. B. Littlewood, Phys. Rev. B {\bf 46}, 405
(1992); L. Coffey and D. Coffey, Phys. Rev. B {\bf 48}, 4184 (1993).

\bibitem{r1} R. Fehrenbacher and M. R. Norman, Phys. Rev. B {\bf 50},
3495 (1994).

\bibitem{r2} R. Fehrenbacher and M. R. Norman, unpublished.

\bibitem{chak} S. Chakravarty, A. Subdo, P. W. Anderson, and S. Strong,
Science {\bf 261}, 337 (1993); A. Sudbo and S. P. Strong, Phys. Rev. B
{\bf 51}, 1338 (1995).

\bibitem{alex}A. A. Abrikosov, Phys. Rev. B {\bf 51}, 11955 (1995)
and Physica C {\bf 244}, 243 (1995).

\end{references}
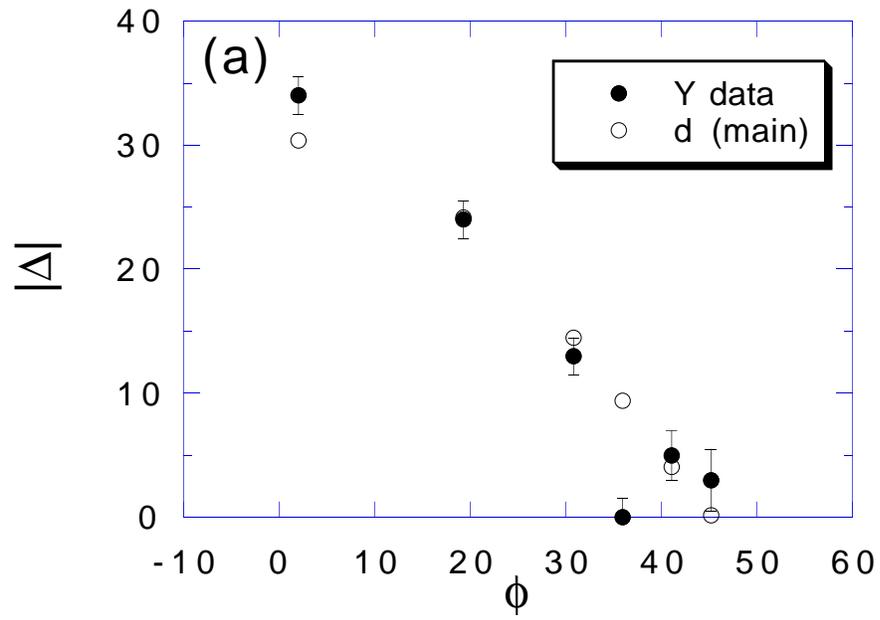
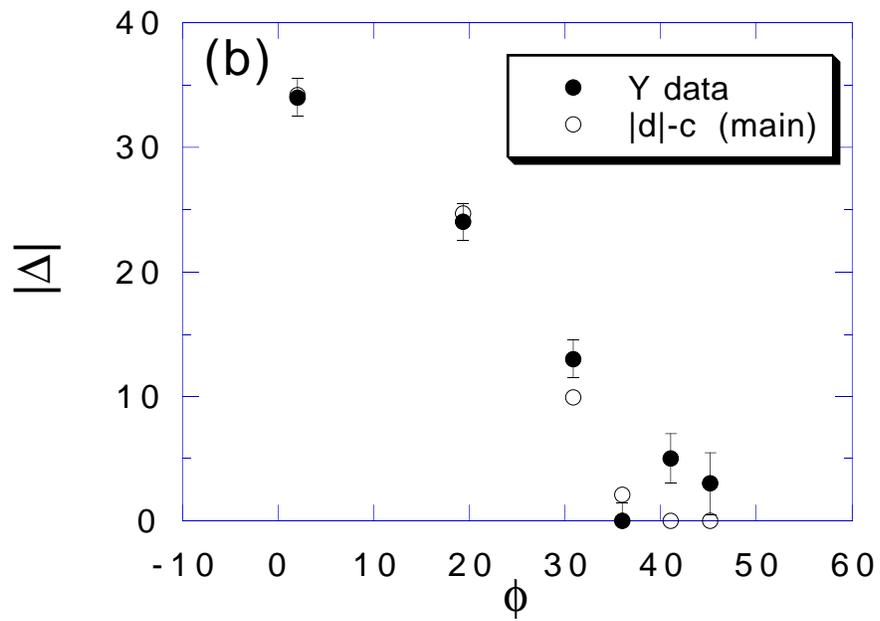
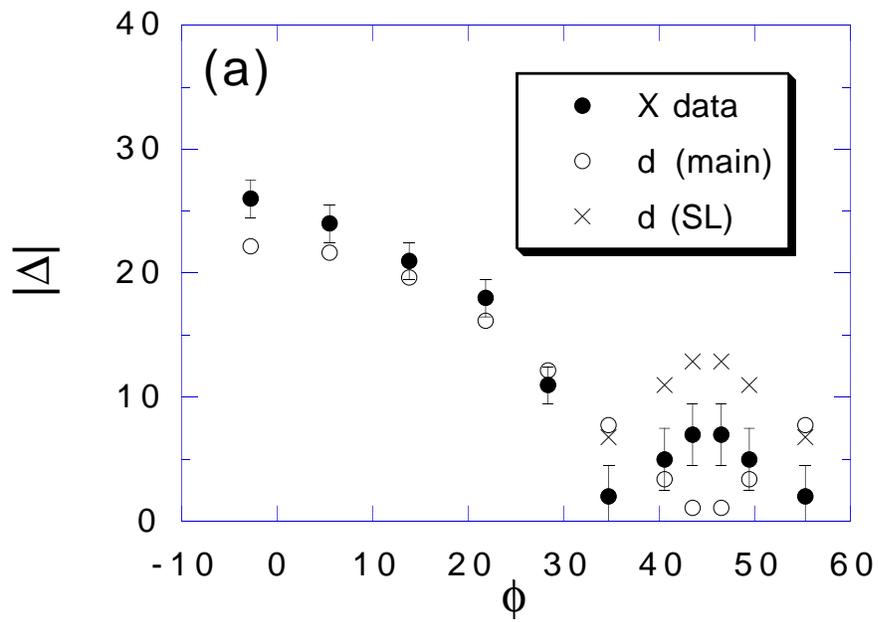
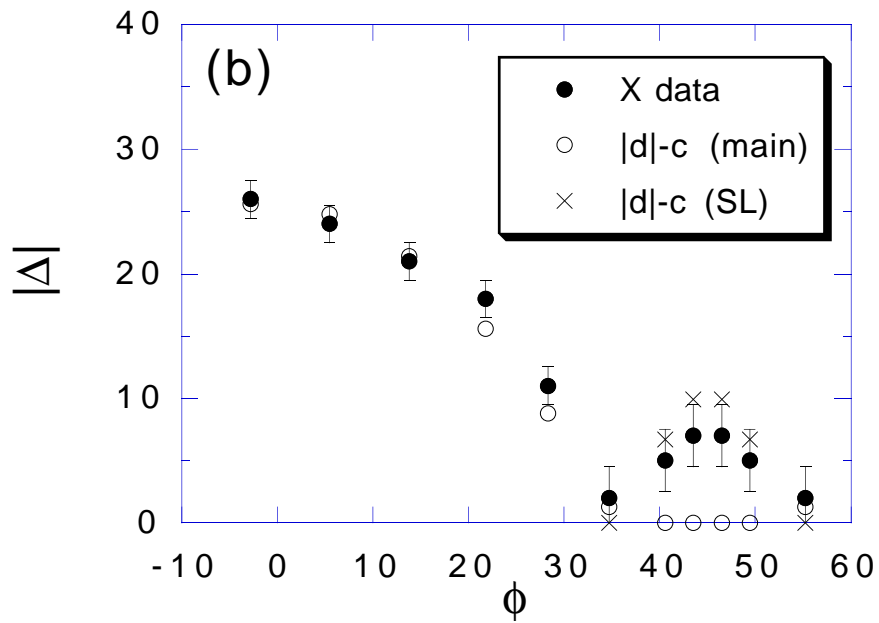
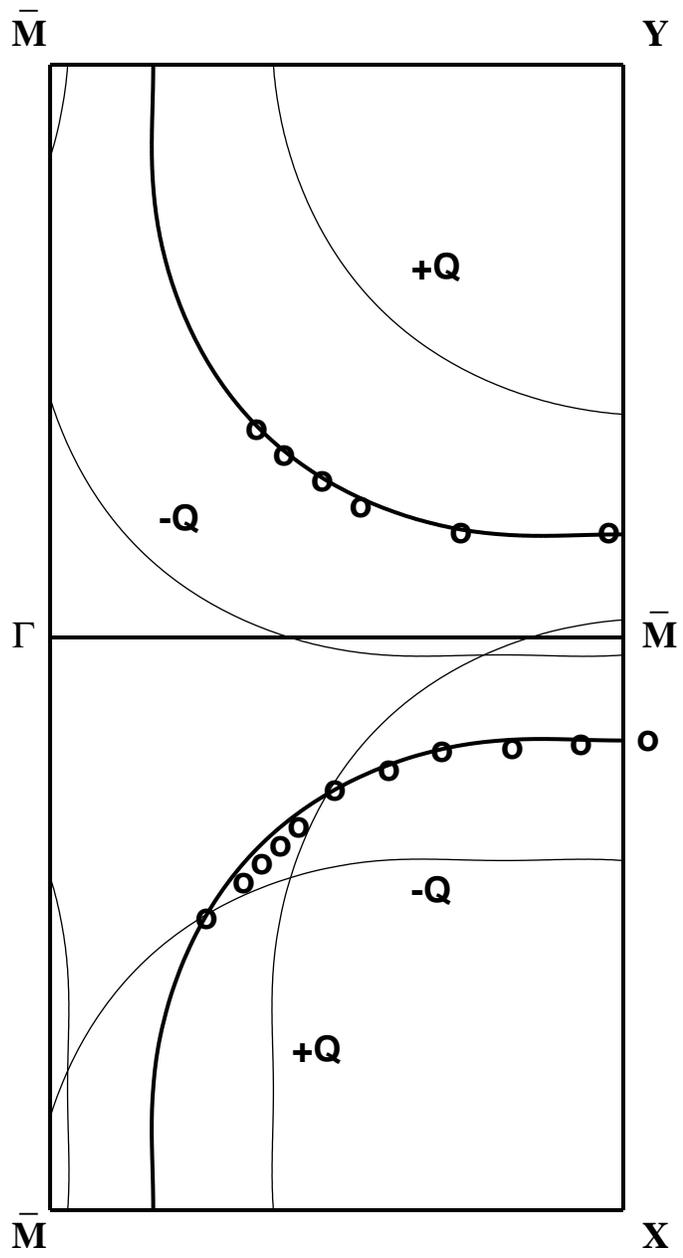
\end{document}